\documentclass[preprint]{aastex}
\usepackage{times}
\usepackage{graphicx}
\IfFileExists{url.sty}{\usepackage{url}}
                      {\newcommand{\url}{\texttt}}

\shorttitle{LLR Eclipse Results}
\shortauthors{Murphy et al.}

\begin{document}

\title{Lunar Eclipse Observations Reveal Anomalous Thermal Performance of
Apollo Reflectors}

\author{
T.\,W. Murphy,~Jr.\altaffilmark{1},
R.\,J. McMillan\altaffilmark{2},
N.\,H. Johnson\altaffilmark{1},
S.\,D. Goodrow\altaffilmark{1}}
\email{tmurphy@physics.ucsd.edu}

\altaffiltext{1}{University of California, San Diego, Dept. of Physics, 9500 Gilman Dr., La Jolla, CA 92093-0424}
\altaffiltext{2}{Apache Point Observatory, 2001 Apache Pt. Rd., Sunspot, NM 88349-0059}

\begin{abstract}
Laser ranging measurements during the total lunar eclipse on 2010 December
21 verify previously suspected thermal lensing in the retroreflectors left
on the lunar surface by the Apollo astronauts. Signal levels during the
eclipse far exceeded those historically seen at full moon, and varied over
an order of magnitude as the eclipse progressed. These variations can be
understood via a straightforward thermal scenario involving solar
absorption by a $\sim 50$\% covering of dust that has accumulated on the front
surfaces of the reflectors. The same mechanism can explain the long-term
degradation of signal from the reflectors as well as the acute signal
deficit observed near full moon.
\end{abstract}

\keywords{Moon, Surface; Instrumentation; Experimental Techniques}

\section{Introduction}

Corner-cube reflectors (CCRs) were placed on the Moon by the Apollo
astronauts during the Apollo~11, Apollo~14, and Apollo~15 landings.
Each reflector consists of an array of solid, circularly-cut fused
silica CCRs 3.8~cm in diameter, installed for the purpose of lunar laser ranging
(LLR) operations that could test gravitational physics, elucidate details
of the lunar interior, and improve knowledge of Earth orientation
and precession \citep{llr-review}.

\begin{figure}
\begin{center}\includegraphics[width=88mm]{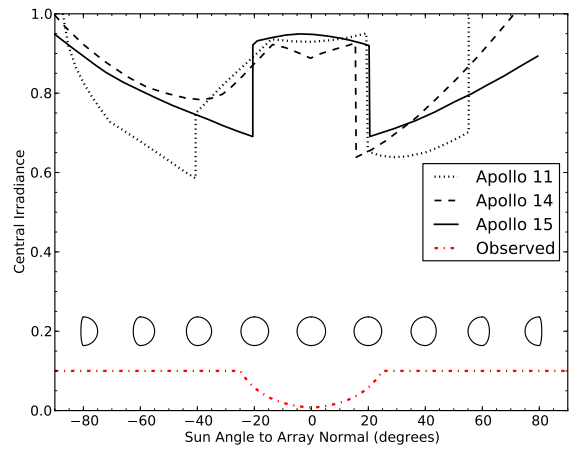}	
\end{center}
\caption{(color online) Expected design performance of the three Apollo reflectors as a
function of sun angle, considering azimuthal orientation, breakthrough of
TIR, and tilt angle of the mounting tray to the lunar surface
\citep{a15-report}.  Adding 180$^{\circ}$ to the horizontal axis
effectively corresponds to the lunar phase angle, $D$, pictorially
represented as illuminated portions of the lunar disk. The dash-dot line
near the bottom represents the approximate best performance observed from
Apache Point in recent years, suffering an overall factor of ten
degradation at most phases and approaching losses in excess of 99\% near full
moon \citep{dust}.\label{fig:LRRR}}
\end{figure}

Soon after commencing LLR observations with the Apache Point Observatory
Lunar Laser-ranging Operation \citep[APOLLO:][]{apollo} in 2006, two
problems became evident. First, the signal strength returning
from the lunar reflectors is diminished by approximately a factor of ten
compared to carefully calculated theoretical expectations
\citep{10x-redux}.  Second, the reflector arrays suffer
an \emph{additional} order-of-magnitude signal reduction when the the lunar
phase is within about 20$^{\circ}$ of full moon \citep{dust}.  Historical
data indicate that the full-moon deficit condition slowly developed
during the first decade after placement on the lunar surface.  The combined
effect of the two facets of signal reduction is that signal strength is
never greater than about 10\% of expectations at any lunar phase, reducing
to $\sim 1$\% near full moon---schematically depicted by the dash-dot line
in Fig.~\ref{fig:LRRR}.

The Apollo CCR arrays were designed and built in an impressive six-month
period by Arthur D. Little, Inc., including a substantial effort dedicated
to thermal design in order to minimize thermal gradients within the
solid prisms. It is well-understood that thermal gradients within
an optical device impose variations in the refractive index, leading
to thermal lensing effects. The central intensity of the far-field
diffraction pattern (FFDP) emerging from the CCR is severely diminished
when differences of even a few degrees Kelvin exist across the corner
cube \citep{ccr-thermal}. Total internal reflection (TIR) corner cubes,
despite producing lower central irradiances compared to CCRs with
reflective coatings, were selected for the Apollo reflectors so that
incoming sunlight would be completely reflected when arriving within
17$^{\circ}$ of normal incidence---and larger incidence angles at
certain azimuth angles. Total reflection of incident energy, and especially
the lack of direct absorption in rear-surface coatings, translates
to reduced thermal gradients within the CCR material. Engineering
documents presented a number of thermal modeling predictions for the
performance, based on the FFDP central irradiance of the reflector
array as a function of sun angle \citep{adl,a15-report}. Incorporating
details of azimuthal orientation and tilt of the reflector tray on
the Moon, the central irradiance for each Apollo reflector was expected
to remain above 60\% of the nominal value for all sun illumination
angles (Fig.~\ref{fig:LRRR}).

Note that around full-moon phase, when the tilted arrays are all facing the
Sun, the reflectors are expected to behave quite well, since this is the
domain in which TIR rejection of incident solar energy is indeed total. We
have found, in contrast, that reflector signal strength is at its worst
near full moon, as indicated by the dash-dot line in Fig.~\ref{fig:LRRR}.

Various possible mechanisms were presented in \citet{dust} to account for
both facets of observed signal reduction simultaneously. Each of the
scenarios involved anomalous absorption or scattering of photons, leading
to both the overall signal deficit and poor performance at full moon via
solar--thermal lensing. The simplest and most plausible
of the scenarios is the slow accumulation of dust on the front surface of
the reflectors as dust is transported across the lunar surface both
electrostatically and by impact activity
\citep{stubbs-dust,farrel-dust,grun-dust}.  \citet{grain-size} found that
intermediate-sized grains approximately 10~$\mu$m in diameter are most
successfully lofted in a simulated space environment.  For grains in this
size range, geometrical obscuration would dominate over diffraction effects
for visible light.

Part of the rationale for attributing the full-moon deficit to a thermal
problem is due to strong performance during past total lunar
eclipses---generally becoming visible within minutes of totality---as
gleaned from the archive of lunar laser ranging normal points available
through the International Laser Ranging Service \citep{ilrs}. This
observation strongly suggests that solar illumination is a key factor.
Because the APOLLO LLR facility is capable of operating in the
high-background conditions at full moon, we had the opportunity to follow
the performance of the reflectors through an entire eclipse event on 2010
December 21.

We present here the heuristic performance expectations of a reflector
array suffering solar-induced thermal gradients during the course
of a total eclipse, exploring briefly the dust deposition that would
be necessary to create the previously reported performance deficits.
We then present the observed performance during eclipse, demonstrating
a close match to the heuristic expectations. We conclude that the
lunar reflectors are not operating according to their design, likely
burdened with a fine layer of dust. Detailed thermal simulations of
the CCRs and mounting trays in the lunar environment are not within
the scope of this paper, for which the primary objective is presentation
of the eclipse observations.

\section{Thermal Expectations}

\begin{figure}
\begin{center}\includegraphics[height=65mm]{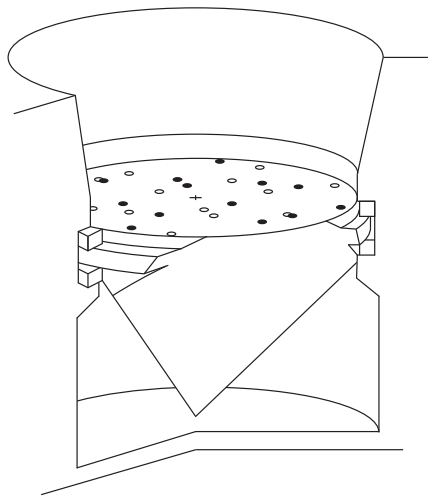}	
\end{center}
\caption{Schematic representation of an Apollo corner cube prism situated
in its aluminum cup (cut-away in drawing), held by Teflon rings sandwiching
the tabs protruding from the prism edges.  Dust grains are illustrated as
dark spots on the front surface of the CCR.  Each real grain has a virtual
analog (open symbols, diametrically opposite the center mark from the real
grain) demarking the entry point for a ray that will ultimately strike the
real grain on exiting the corner cube.  The covering fraction represented
in the drawing is substantially less than that posited in this paper, where
real (filled) grains obscure approximately half of the surface area.
\label{fig:CCR-mount}}
\end{figure}

We have detailed separately the effect of axial and radial thermal
gradients within a CCR on the central irradiance of the FFDP. The
conclusion is that a temperature difference across the CCR of only a few
degrees can destroy the central irradiance \citep{ccr-thermal}.  A simple
model for what may be plaguing the lunar reflectors is that dust on the
front surface absorbs solar radiation when the array points nearly face-on
to the Sun---as is the case near full moon. The CCRs are recessed into an
aluminum tray by half their diameters (see Fig.~\ref{fig:CCR-mount}), so
that illumination of the front surface is complete only at full phase.
Solar energy absorbed by the dust is radiatively and conductively
transferred into the front surface of the CCR, creating a thermal gradient
within the CCR that was not anticipated in the design. The gradient
translates to a varying refractive index, or thermal lensing, imparting
phase delays for different optical paths within the CCR.

For instance, if the front surface of the corner cube is hotter than its
vertex, a ray path entering and exiting the CCR near its outer radius will
stay relatively close to the front surface as it traverses the interior of
the CCR, experiencing a slightly larger average refractive index and
therefore greater phase delay compared to a central ray that penetrates
deep into the CCR and into cooler material. The result is a spherical
wavefront advanced in the center and retarded at the edges. The divergence
translates into a loss of peak intensity in the far field, and thus reduced
return signal. Radial temperature gradients produce similar-scale effects
on the wavefront and FFDP. Thermal expansion also plays a role, but far
less pronounced than the refractive effect \citep{ccr-thermal}.

\subsection{Dust Covering Fraction and Thermal Impact\label{sec:thermal-model}}

We model the putative front-surface dust absorption as covering a
fraction, $f$, of the front surface in small grains randomly and uniformly
distributed across the surface.  Assigning to the dust an albedo,
$\alpha\approx0.1$, results in a front-side thermal input in full sun of
$I_{0}A(1-\alpha)f$, where $I_{0}\approx1370$~W~m$^{-2}$ is the solar
irradiance and $A$ is the frontal area. Light that successfully enters the
CCR will re-emerge through the front surface after retroreflection, to
again find probability, $f$, of absorption by dust (see
Fig.~\ref{fig:CCR-mount} for a schematic example). The probability of
transmission through both passages of the front surface is $(1-f)^{2}$,
assuming random (independent) placement of dust grains on the surface.  The
total rate of energy absorbed by dust for the Apollo CCRs then computes
to:\begin{equation}
P_{\mathrm{absorb}}=I_{0}A(1-\alpha)f(2-f)\approx1.41(2f-f^{2})\,\mathrm{W}.\label{eq:absorbed}\end{equation}

In order to explain the observed order-of-magnitude signal deficit
at all phases, we make use of the fact that the double-pass of the
CCR front surface results in an effective likelihood of transmission
proportional to $(1-f)^{2}$, reducing the effective aperture area
by this same factor. The central irradiance of the far field diffraction
pattern arising from an arbitrary aperture is: \begin{equation}
I_{\mathrm{central}}=\left|\int\int_{\mathrm{aperture}}E(u,v)\exp[i\phi(u,v)]\mathrm{d}u\mathrm{d}v\right|^{2},\label{eq:cent-irrad}\end{equation}
where $E(u,v)$ is the electric field at position $(u,v)$ in the
aperture plane, and $\phi(u,v)$ is the associated phase. For a plane
wave of uniform intensity, we see that the central irradiance is simply
proportional to the square of the aperture area---independent of configuration.
The effective area of a dust-covered CCR will be $(1-f)^{2}$ times
the geometrical area, so that the return strength in the far-field
should scale as $(1-f)^{4}$. From our observation that the CCR signal
is reduced by a factor of ten or more (a factor of 15 may be closer
to the truth), we put the fill-factor, $f\sim0.5$, so that the front-surface
thermal absorption from Eq.~\ref{eq:absorbed} becomes approximately
1.0~W.

The thermal power radiated to space from the front surface of each CCR depends
on the temperature the CCRs reach under full sun, which in part depends
on the temperature attained by the mounting tray. This, too, depends
in principle on the state of dust coverage. A perfect blackbody equilibrates
at 394~K under 1370~W~m$^{-2}$ of solar illumination, as does
lunar dust, with ``balanced'' values for albedo ($\alpha\approx0.1$)
and emissivity ($\varepsilon=0.9$). In their thermal models, the
manufacturers of the reflector arrays used 0.2 for both optical absorptance
($1-\alpha$) and emissivity for the aluminum top surface of the array,
the balance of which also results in an equilibrium temperature of
394~K. It is worth noting that preparation of the aluminum surface
can have a dramatic impact on equilibrium temperature: machined aluminum
($\alpha\approx0.9$, $\varepsilon\approx0.05$) reaches 470~K in
full sunlight, while clear-anodized aluminum ($\alpha\approx0.85$,
$\varepsilon\approx0.8$) is a far cooler 260~K. A 50\% lunar dust
covering on said anodized aluminum would raise the temperature to
330~K.

Adopting 394~K as an upper limit to the CCR temperature in full sun,
the front-surface ($\varepsilon\approx0.9$) radiated power would
be as large as $-1.4$~W if the radiation were able to couple into
$2\pi$~sr of cold sky. However, the Apollo CCRs are recessed by
half their diameters into slightly conical clear-anodized aluminum
pockets (1.5$^{\circ}$ half-angle for Apollo~11; 6$^{\circ}$ for
the other two; see Fig.~\ref{fig:CCR-mount}). This results in area-weighted solid angles to cold
space of 1.52 and 1.70~sr, for the two designs, or only 24 and 27\%
of a $2\pi$~sr half-space. Since the conical walls are known to
have moderately high emissivity \citep[$\varepsilon\approx0.8$;][]{adl},
the radiative cooling power of the CCR face into space is reduced
to $\sim-0.6$~W---assuming a fused silica emissivity of $\varepsilon\approx0.9$
and that any radiation not absorbed by the conical walls is reflected
to space. Note that this is less than the estimated $\sim1.0$~W
rate of solar absorption by dust on the front surface. Dust on the
conical walls does little to modify the radiative cooling rate, as
the walls are already high-emissivity. But if anything, the presence
of dust would reduce cooling efficacy further.

We may therefore expect the CCR front surface to see a net influx
of thermal power, the associated energy making its way to the aluminum
mounting tray by radiation (suppressed by the low emissivity cavity
behind the CCR) and mount conductance through the tabs and Teflon
rings around the periphery of the CCR (Fig.~\ref{fig:CCR-mount}). In principle, the relative
efficacies of these two flows will determine whether axial or radial
gradients dominate. The distinction is not important for the purposes
of this paper, as either type of gradient diminishes the far-field
signal strength \citep{ccr-thermal}. A detailed thermal simulation
including the mounting tray, the lunar surface, and the impact of
dust would be desirable to confirm the scenario described here, but
is beyond the scope of the current paper.

\begin{figure}
\begin{center}\includegraphics[width=88mm]{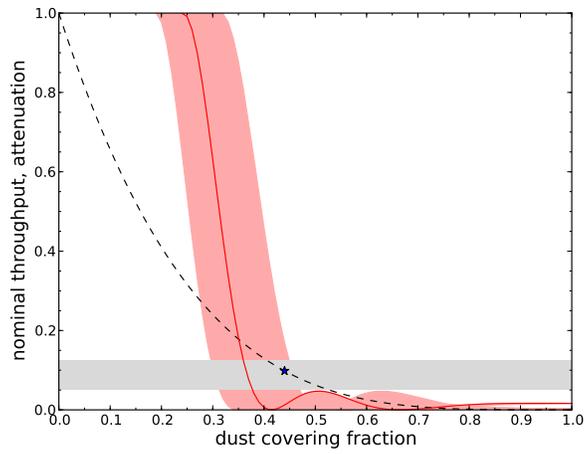}	
\end{center}
\caption{(color online) Schematic constraints on dust covering fraction as relating to
overall throughput and thermal lensing.  The horizontal gray band indicates
the approximate factor of ten by which both dust obscuration and the full
moon deficit are observed to operate.  The dashed line captures the
$(1-f)^4$ direct obscuration effect, by itself suggesting a covering
fraction around 0.4--0.5.  The solid line reflects the thermally-induced
attenuation expected in full sunlight, based on a front-surface radiative
cooling rate of $-0.6$~W, and surrounded by a band accommodating 25\%
uncertainty in this rate.  Both obscuration and thermal lensing can be
satisfied for a dust covering fraction around 0.40--0.46, as indicated by
the star.  The thermal curve is based on an axial gradient, so is only
suggestive of a realistic thermal distribution.
\label{fig:frac-dep}}
\end{figure}

Short of a full thermal simulation, we can at least investigate the
plausibility of optically significant thermal gradients based on the
fact that the net power, $P$, transmitted through the CCR front surface
will produce a temperature gradient in the neighborhood of $\Delta T\approx Pd/\kappa A$,
where $d$ is the ``thickness'' across which the gradient develops,
and $\kappa=1.38$~W~m$^{-1}$~K$^{-1}$ is the thermal conductivity
of fused silica. In order to realize debilitating thermal gradients
\citep[of order 4~K; see][]{ccr-thermal}, we require a front-surface
power imbalance of $\sim0.3$~W, using the CCR radius as the characteristic
``thickness,'' $d$. This is comparable to the difference calculated
above of a 1.0~W input in conjunction with a $-0.6$~W radiative loss
rate. These numbers are not precisely determined, but set a scale
that is fully consistent with gradients several degrees in magnitude,
and thus optically important.  Fig.~\ref{fig:frac-dep} illustrates the
feasibility of both obscuration and thermal lensing imposing the
observed factor-of-ten signal deficit effects at the same dust covering
fraction.

\subsection{Thermal Time Constant}

The time constant for changing a thermal gradient in a solid mass
of fused silica can be approximated by evaluating the timescale over
which an axial gradient in a cylinder of material having height $h$
and radius $R$ would equilibrate. At a thermal conductivity $\kappa$,
the rate of energy flow along the cylinder axis will be $\kappa\pi R^{2}\Delta T/h$,
where $\Delta T$ is the total temperature difference across the cylinder.
If the temperature is allowed to equilibrate with no flow of heat
into or out of the cylinder, we must move $c_{\mathrm{p}}\rho\pi R^{2}h\Delta T/8$
of thermal energy from one half of the cylinder to the other half.
Here, $c_{\mathrm{p}}$ is the specific heat capacity of the material,
$\rho$ is the density, and the factor of eight transforms the end-to-end
temperature difference into the amount of excess thermal energy that
must be moved elsewhere. Dividing energy transferred by the initial
rate of energy flow yields a time constant\begin{equation}
\tau\sim\frac{c_{\mathrm{p}}\rho h^{2}}{8\kappa}.\label{eq:tau-cyl}\end{equation}
For a fused silica CCR with a height $h=\sqrt{2}R$ ($\rho=2201$~kg~m$^{-3}$;
$c_{\mathrm{p}}=703$~J~kg$^{-1}$K$^{-1}$; $\kappa=1.38$~W~m$^{-1}$K$^{-1}$;
$R=19$~mm for the Apollo CCRs\footnote{materials values from \texttt{http://www.sciner.com/Opticsland/FS.htm}}),
the cylindrical timescale computes to 100 seconds. While weak coupling
to the mounting tray may lengthen the time constant somewhat, one
may still expect to see changes in the thermal state of the CCR well
within the hours-long timescale of the eclipse, given the foregoing
estimation.

\subsection{Predicted Behavior\label{sec:prediction}}

\begin{figure}
\begin{center}\includegraphics[width=88mm]{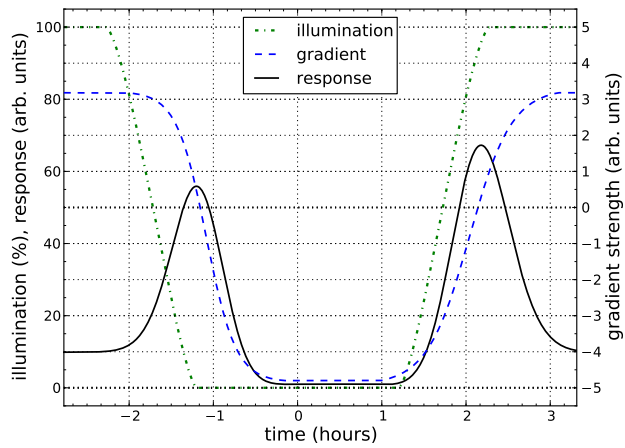}	
\end{center}
\caption{(color online) Cartoon representation of a possible reflector
response scenario during eclipse. Solar illumination is represented by the
dash-dot curve, and refers to the left axis. Prior to the eclipse, the
gradient (dashed curve, right axis) is positive due to solar illumination,
swinging (potentially more) negative during the radiative cooling phase.
The response (solid curve, left axis) peaks when the gradient crosses
through zero, although the speed with which it does so need not be the same
for ingress and egress. The peak response is likely sensitive to the
rapidity of the gradient change as it crosses through zero, reflected in
the cartoon. Timescales for signal changes are merely suggestive, and do
not represent a thermal simulation. Heavy dotted lines indicate zero for
the left and right axes. \label{fig:cartoon}}
\end{figure}

The eclipse provides a celestial light switch with which to test our
hypothesized thermal scenario by tracking the reflected signal strength
as the solar illumination changes. Operating under the assumption
that our model for solar absorption by dust at the front surface is
correct, the front of the CCR under solar illumination will be hotter
than the bulk, establishing a positive thermal gradient (we will adopt
the convention that the gradient is positive when the front face is
hotter than the interior bulk). During the eclipse, we can expect
the CCR to radiate its stored thermal energy to space via the high-emissivity
front surface, ultimately resulting in the establishment of a negative
gradient. At some point during this reversal, the overall gradient
should cross through zero---at which time we should expect a strong
return due to an approximately isothermal corner cube prism. Depending
on the magnitude of the negative gradient that develops well into
the eclipse, the return may become weaker than the normal full-moon
performance. When sunlight reappears, the gradient again reverses,
passing through zero once more as the CCR transitions from a negative
to a positive thermal gradient. Therefore, we may expect a second
peak in return strength around this time, before settling back to
the usual full-moon performance under the influence of a positive
gradient. Fig.~\ref{fig:cartoon} illustrates a possible evolution
during the eclipse.

Because the eclipse-induced thermal transitions happen quickly, a
truly isothermal condition is unlikely to develop during the process.
But the dominant gradient should be largely ameliorated during the
reversal. The speed with which the gradient changes may influence
the degree of thermal anisotropy, and hence the peak return strength.
Because the net heating or cooling rate of the CCR is not likely to
be symmetric during ingress and egress---this depends on the detailed
balance of front-surface absorption and radiation---we should not
be surprised to see a different evolution of return strength during
egress vs. ingress. Such an asymmetry is suggestively illustrated
in Fig.~\ref{fig:cartoon}. For example, using the estimated rates
from Section~\ref{sec:thermal-model}, the net heat flow into the
corner cube before the eclipse will be 0.4~W, while during eclipse
the rate will be $-0.6$~W (only radiation). In this scenario, one
would expect the stabilized negative gradient to exceed the stabilized
positive gradient in absolute value, and a faster transition accompanying
eclipse ingress than that for egress, as depicted in Fig.~\ref{fig:cartoon}.

\section{Observations and Results}

\subsection{Observing Conditions and Timing}

The total lunar eclipse on 2010 December 21 (near solstice) was well-placed
in the sky at the Apache Point Observatory, reaching an elevation
angle of $\sim80^{\circ}$ around the time that totality began. High,
thin clouds of variable thickness were present throughout the observations.
Moderate and fairly steady winds were also present throughout the
period: one minute samples indicate a mean, median, and standard deviation
of 37.2, 38.6, and 4.7~km~hr$^{-1}$, respectively, having a minimum
and maximum of 20.9 and 48.3~km~hr$^{-1}$. Surface wind direction
was also steady, staying between azimuths of 255$^{\circ}$ and 280$^{\circ}$
(originating from the west). The impact of wind on tracking increased
during the night, as the telescope and open dome shutter followed
the Moon into the west.

\begin{figure}
\includegraphics[height=75mm]{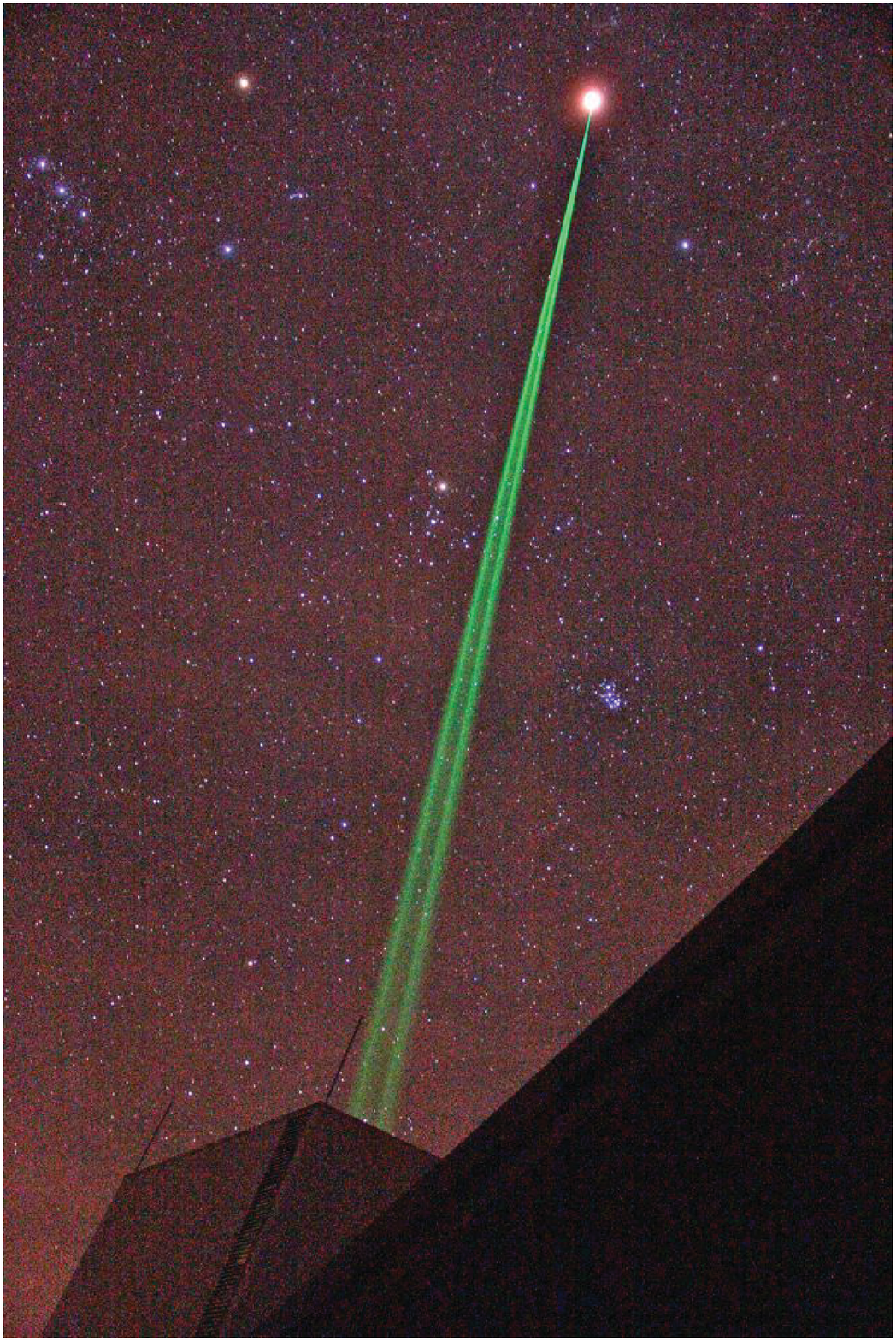}	
\hfill{}\includegraphics[height=75mm]{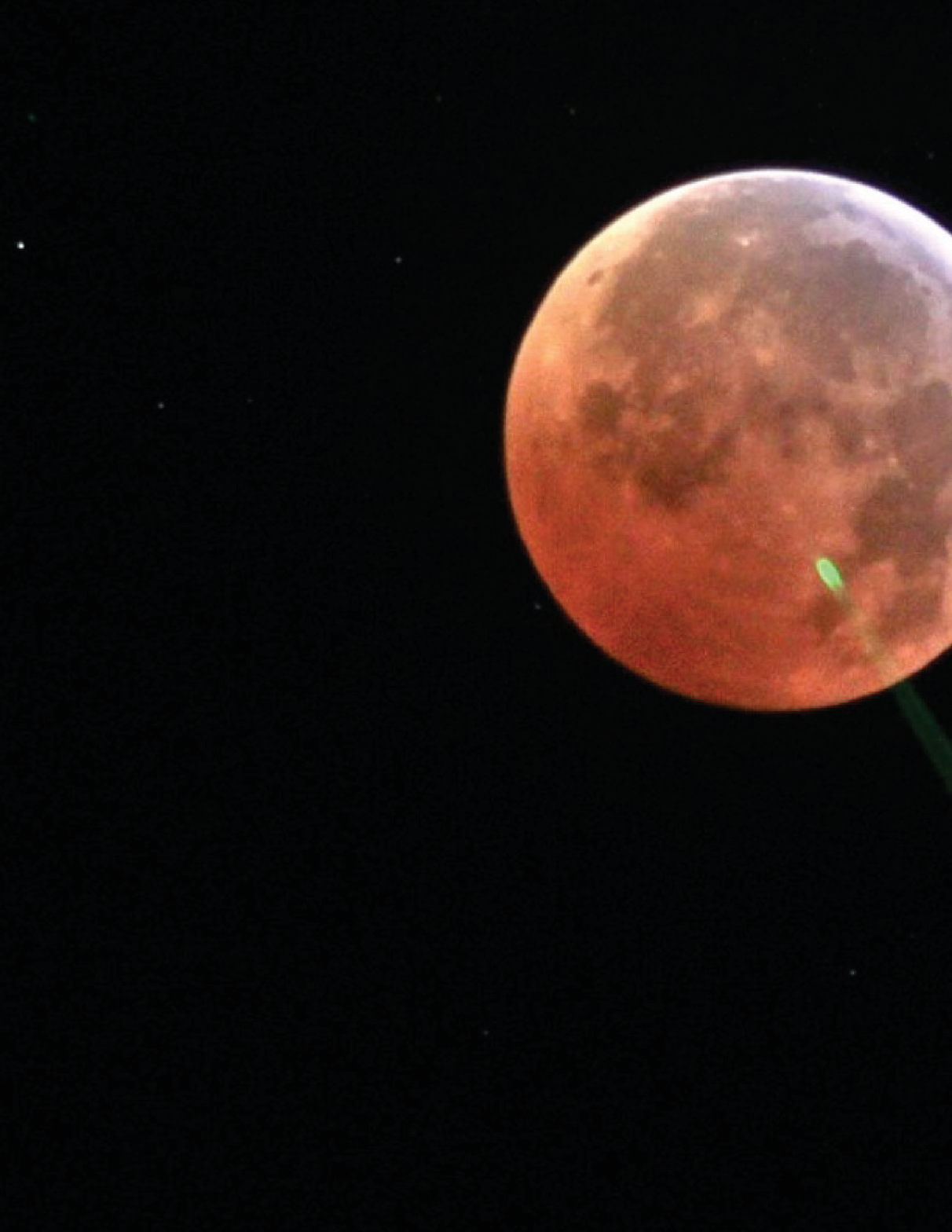}	
\caption{(color online) Photographs taken during the eclipse showing (at
left; taken at 07:53:00~UTC) the overall clarity of the sky as the laser
beam crosses in front of Taurus, although very thin clouds were present at
high altitude. The laser spot in the right-hand image (taken at
08:22:45~UTC) is reflected by said clouds and not by the lunar surface: the
target at the time was Apollo~14, located just left of center on the lunar
disk. Photos by Jack Dembicky.\label{fig:photos}}
\end{figure}

The timeline for the eclipse appears in Table~\ref{tab:eclipse-timeline}.
Laser ranging attempts commenced at 05:50~UTC, ending at 11:20~UTC,
when the Moon was at elevations of $73^{\circ}$ and $34^{\circ}$,
respectively. Fig.~\ref{fig:photos} shows pictures taken during
the total phase of the eclipse from a location approximately 40~m
from the telescope. The spot of laser light visible at lower right
in the right-hand image is reflected from thin clouds about 10~km
high. The laser power measured 1.29--1.42~W during the observation
period---less than the long-term average nearer 2~W.

\begin{table}

\caption{Eclipse Timeline\label{tab:eclipse-timeline}}

\begin{center}\begin{tabular}{lc}
\hline 
Event&
Time (UTC)\\
\hline
P1: Penumbral eclipse begins (first contact)&
05:29:17\\
U1: Partial eclipse begins (second contact)&
06:32:37\\
U2: Total eclipse begins (third contact)&
07:40:47\\
Eclipse midpoint&
08:16:57\\
U3: Total eclipse ends (fourth contact)&
08:53:08\\
U4: Partial eclipse ends (fifth contact)&
10:01:20\\
P2: Penumbral eclipse ends (sixth contact)&
11:04:31\\
\hline
\end{tabular}\end{center}
\end{table}

\begin{figure}
\begin{center}\includegraphics[width=88mm]{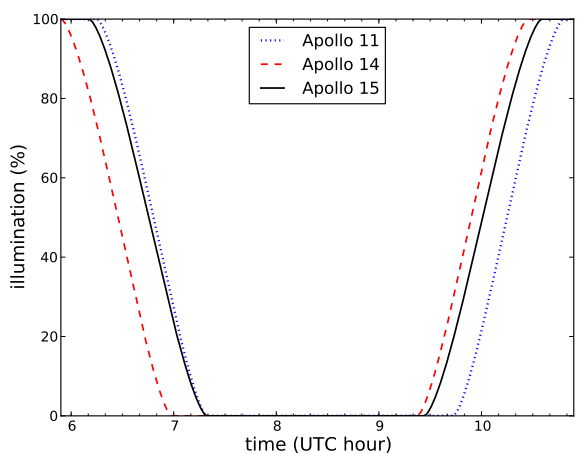}	
\end{center}
\caption{(color online) Illumination history of the three Apollo reflectors
on 2010 December 21. Note that the eclipse duration for Apollo~15 (black)
is shorter than for the other two reflectors. Note also that Apollo~14
(red) and Apollo~11 (blue) follow parallel paths due to their similar
selenographic latitudes and the nearly pure right-ascension trajectory of
the moon at the time. \label{fig:illum-curves}}
\end{figure}

Because the edge of the Earth's shadow is curved, and the transit
of the eclipse through the shadow was not centered, different reflectors
not only see time offsets in the shadow crossing, but different durations
and onset slopes. As a result, a different illumination timeline applies
to each reflector during the eclipse, as illustrated in Fig.~\ref{fig:illum-curves}. 

\begin{figure}
\includegraphics[width=88mm]{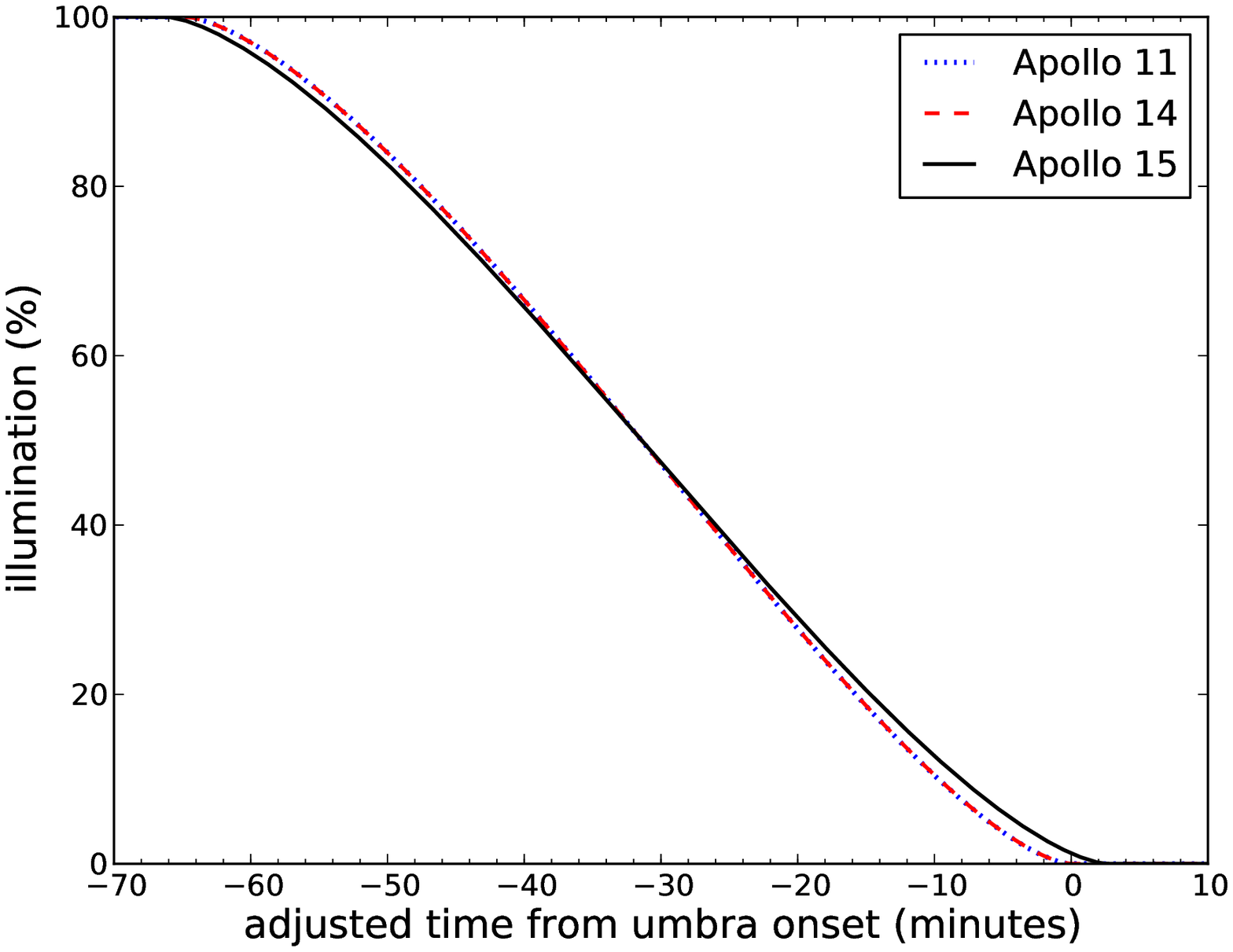}	
\hfill{}\includegraphics[width=88mm]{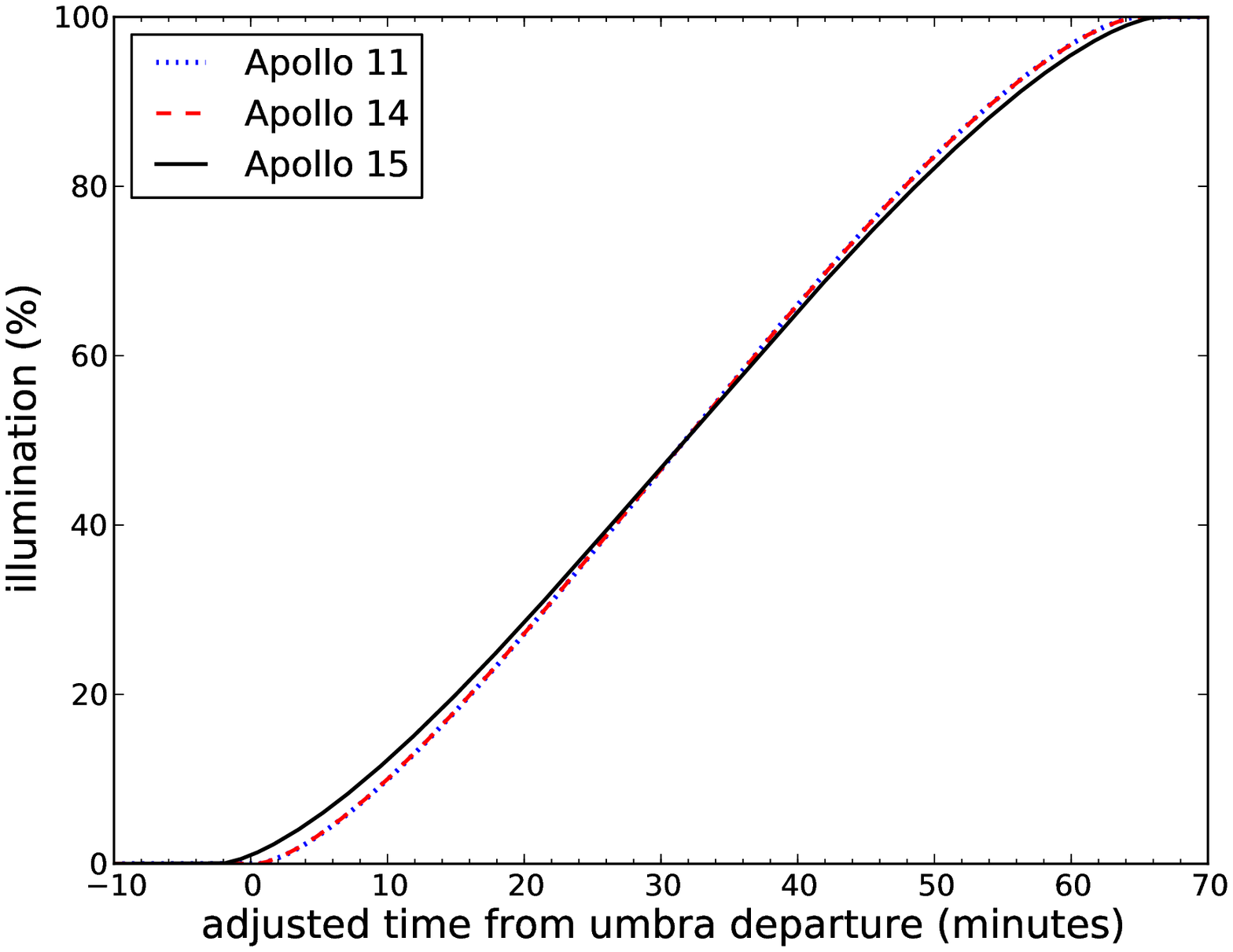}	
\caption{(color online) Ingress (left) and egress (right) curves, aligned
by time adjustments (offsets) to overlap at the 50\% illumination mark.
\label{fig:ingress-egress}}
\end{figure}

It is to our advantage to evaluate the temporal response of the reflectors
with respect to their individual illumination histories, rather than
on an absolute time axis. If we shift the illumination history curves
in time so that the midpoint of both ingress and egress separately
align at the 50\% illumination points, we find that we must shift
the ingress time of Apollo~11 and Apollo~15 earlier by 20.8 and
17.9 minutes, respectively, to lie on top of the (earliest) Apollo~14
curve. Likewise, for egress, Apollo~11 and Apollo~15 must be shifted
earlier by 20.9 and 6.7 minutes, respectively. Fig.~\ref{fig:ingress-egress}
shows the aligned curves---which for Apollo~11 and Apollo~14 lie
on top of each other, and are indistinguishable.

\subsection{Data Acquisition and Presentation}

\begin{figure}
\begin{center}\includegraphics[width=88mm]{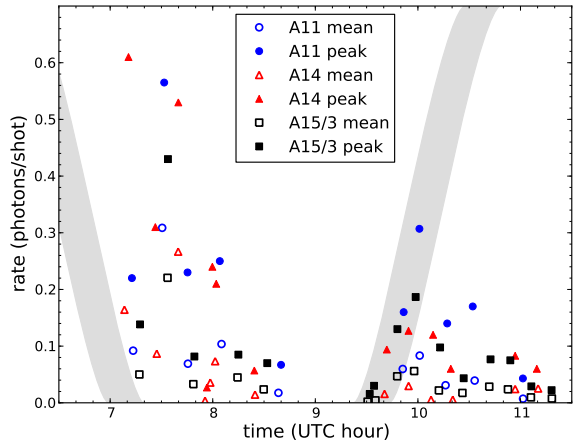}	
\end{center}
\caption{(color online) Observed return rates as a function of the actual
time of observation. Both mean photon rates and peak photon rates are
shown. The Apollo~15 rates have been reduced by a factor of three to allow
direct comparison to the smaller Apollo~11 and Apollo~14 reflectors. Times
correspond to either the midpoint of the run or the time of the peak for
mean rates and peak rates, respectively.  The light gray band indicates the
range of illumination curves for the three Apollo reflectors, as shown in
Fig.~\ref{fig:illum-curves}.  \label{fig:raw-rates}}
\end{figure}

The raw observations are detailed in Tables~\ref{tab:ingress} and
\ref{tab:egress}, and shown in Fig.~\ref{fig:raw-rates}. For each
run, whose start time is listed in the table and spans 50~s per 1000
shots, a reflector signal is detected against background events, summed
in a 1~ns temporal window following the predicted lunar range
\citep[see][for example data]{apollo}. The background rate is subtracted,
and an average rate computed, representing the number of reflector
photons divided by the number of laser shots. Before plotting results,
rates for Apollo~15 are divided by 3 to account for its larger size
so that the reflector responses may be compared to one another. We
also present a peak rate corresponding to the best 15~s contiguous
period within each run. Comments capture supplemental notes made at
the time of observation; notes about wind do not reflect anomalously
windy periods, according to wind records for the night, although their
influence on pointing likely increased over time as the telescope
tracked into the wind.

\begin{table}

\caption{Observations during ingress\label{tab:ingress}}
\scriptsize
\begin{tabular}{cccccccl}
\hline 
Time&
Reflector&
Illum.&
Shots&
Photons&
Mean&
Peak&
Comments\\
(UTC)&
&
(\%)&
&
&
&
&
\\
\hline
5:50:17&
A15&
100&
10000&
---&
---&
---&
seeing $\approx$1.5 arcsec; laser power 1.35 W\\
5:59:52&
A15&
100&
2477&
---&
---&
---&
cut short for satellite block\\
6:08:32&
A15&
100&
5000&
---&
---&
---&
\\
6:13:03&
A15&
98&
5000&
---&
---&
---&
\\
6:18:33&
A14&
63&
5000&
---&
---&
---&
\\
6:23:29&
A15&
86&
5000&
---&
---&
---&
\\
6:36:12&
A15&
64&
5000&
---&
---&
---&
after pointing update\\
6:41:16&
A14&
20&
5000&
---&
---&
---&
\\
6:46:36&
A15&
43&
5000&
---&
---&
---&
\\
6:59:13&
A15&
19&
5000&
---&
---&
---&
after transmit/receive offset\\
7:05:51&
A14&
0&
6000&
985&
0.164&
0.610&
\\
7:11:46&
A11&
5&
4000&
368&
0.092&
0.220&
\\
7:15:50&
A15&
0&
3000&
450&
0.15&
0.415&
windy $\rightarrow$ highly variable rate\\
7:21:38&
L1&
0&
4500&
---&
---&
---&
\\
7:26:00&
A14&
0&
3000&
260&
0.087&
0.310&
\\
7:29:11&
A11&
0&
3000&
926&
0.309&
0.565&
\\
7:32:24&
A15&
0&
3000&
1985&
0.662&
1.290&
\\
7:35:24&
L1&
0&
3000&
---&
---&
---&
\\
7:38:32&
A14&
0&
3000&
800&
0.267&
0.530&
\\
7:44:13&
A11&
0&
3000&
207&
0.069&
0.230&
\\
7:47:23&
A15&
0&
3000&
296&
0.099&
0.245&
windy $\rightarrow$ highly variable rate\\
7:50:59&
L1&
0&
3000&
---&
---&
---&
\\
7:54:15&
A14&
0&
3000&
11&
0.004&
0.027&
essentially lost (tracking drift?)\\
7:57:19&
A14&
0&
3000&
107&
0.036&
0.240&
\\
8:00:10&
A14&
0&
3000&
220&
0.073&
0.210&
\\
8:03:53&
A11&
0&
3000&
311&
0.104&
0.250&
\\
8:12:24&
A15&
0&
5000&
671&
0.134&
0.255&
\\
8:17:31&
L1&
0&
5000&
---&
---&
---&
\\
8:22:39&
A14&
0&
5000&
72&
0.014&
0.057&
\\
8:27:43&
A15&
0&
5000&
355&
0.071&
0.210&
\\
8:36:20&
A11&
0&
5000&
88&
0.018&
0.067&
\\
8:41:37&
A14&
0&
5000&
---&
---&
---&
IR all-sky camera showing clouds worse\\
8:46:41&
A15&
0&
6000&
---&
---&
---&
\\
8:52:38&
A15&
0&
4000&
---&
---&
---&
laser power 1.42 W after run\\
9:01:00&
A15&
0&
5000&
---&
---&
---&
after run, noted large image motion on camera \\
\hline
\end{tabular}
\end{table}

\begin{table}

\caption{Observations during egress.\label{tab:egress}}
\scriptsize
\begin{tabular}{cccccccl}
\hline 
Time&
Reflector&
Illum.&
Shots&
Photons&
Mean&
Peak&
Comments\\
(UTC)&
&
(\%)&
&
&
&
&
\\
\hline
9:20:02&
A15&
0&
5000&
---&
---&
---&
after waiting for light for crater pointing ref. \\
9:28:30&
A15&
0.5&
5000&
30&
0.006&
0.047&
\\
9:33:06&
A15&
5&
5000&
65&
0.013&
0.090&
\\
9:37:59&
A14&
24&
6000&
94&
0.016&
0.094&
\\
9:46:38&
A15&
25&
3000&
421&
0.140&
0.390&
\\
9:49:53&
A11&
7&
3000&
179&
0.060&
0.160&
\\
9:53:18&
A14&
51&
3000&
89&
0.030&
0.127&
jumpy pointing---presumably wind shake\\
9:56:32&
A15&
44&
3000&
504&
0.168&
0.560&
\\
9:59:48&
A11&
23&
3000&
250&
0.083&
0.307&
\\
10:05:37&
A14&
76&
5000&
26&
0.005&
0.120&
\\
10:10:36&
A15&
72&
4000&
262&
0.066&
0.293&
\\
10:14:31&
A11&
53&
4000&
124&
0.031&
0.140&
high clouds approaching\\
10:18:43&
A14&
95&
4000&
22&
0.006&
0.060&
\\
10:24:22&
A15&
94&
4000&
209&
0.052&
0.130&
\\
10:31:32&
A11&
84&
4000&
157&
0.039&
0.170&
\\
10:35:39&
A14&
100&
4000&
---&
---&
---&
\\
10:40:04&
A15&
100&
4000&
342&
0.086&
0.230&
\\
10:44:16&
A11&
99&
4000&
---&
---&
---&
\\
10:51:00&
A15&
100&
4000&
285&
0.071&
0.225&
\\
10:55:13&
A14&
100&
4000&
97&
0.024&
0.083&
\\
10:59:50&
A11&
100&
4000&
29&
0.007&
0.043&
\\
11:04:26&
A15&
100&
4000&
115&
0.029&
0.087&
\\
11:08:39&
A14&
100&
4000&
101&
0.025&
0.060&
still jumpy pointing (wind shake)\\
11:12:56&
A11&
100&
4000&
---&
---&
---&
\\
11:17:07&
A15&
100&
3000&
68&
0.023&
0.067&
laser power 1.29 W after run\\
\hline
\end{tabular}
\end{table}

Lunar laser ranging is a low-signal endeavor, and some time is inevitably
spent trying to recover a lost signal. Also, the return rate is highly
sensitive to atmospheric conditions---especially turbulence-induced
``seeing''---so that even perfect alignment and tracking produce
factors of 3--5 in signal rate fluctuations on timescales ranging
from seconds to minutes. For these reasons, the reported rates should
all be interpreted with factor-of-two uncertainties. The peak rate
may be more representative of the intrinsic performance than the average
rate, because at least these moments tend to ensure that pointing
and atmospheric transmission are simultaneously as good as they may
be, and therefore more meaningfully compared from one run to another.

During ingress, we began observations at 05:50~UTC, but were not
able to acquire returns from either Apollo~15 or Apollo~14 for the
first hour. At 06:52 we adjusted the transmit-receive beam alignment
offset by 1.6~arcsec: a large change for APOLLO that can easily account
for the lack of signal. After this adjustment, we failed once more
to raise Apollo~15 in a run starting at 06:59. All reflectors were
still illuminated to some degree during these failed attempts. In
the next run, starting at 7:05, we got a very bright return from Apollo~14---about
an order of magnitude stronger than any full-moon result over the
previous five years. We were then able to see Apollo~11 and Apollo~15
prior to their entry into full shadow, displaying comparatively poor
performance, although still better than historical experience. Attempts
were made at 07:21, 07:35, 07:51, and 8:17 to find the Lunokhod~1
reflector, but to no avail. As discussed below, this is not particularly
surprising given the construction of the Lunokhod arrays. 

After 08:41, well into the total phase of the eclipse, we lost the
signal from all reflectors and were unable to recover signal from
Apollo~15 in multiple subsequent attempts. Slightly thicker clouds
during this period may have played a role (explored more thoroughly
later). A run starting at 09:28 recovered a very weak signal from
Apollo~15, just as it was re-entering sunlight. A second peak in
performance was then observed for all reflectors, although less impressive
than the first. Even so, the observed rates were well in excess of
historical return rates at full phase. The last hour of observation
produced signal levels typical of full-moon phase.

\begin{figure}
\includegraphics[width=88mm]{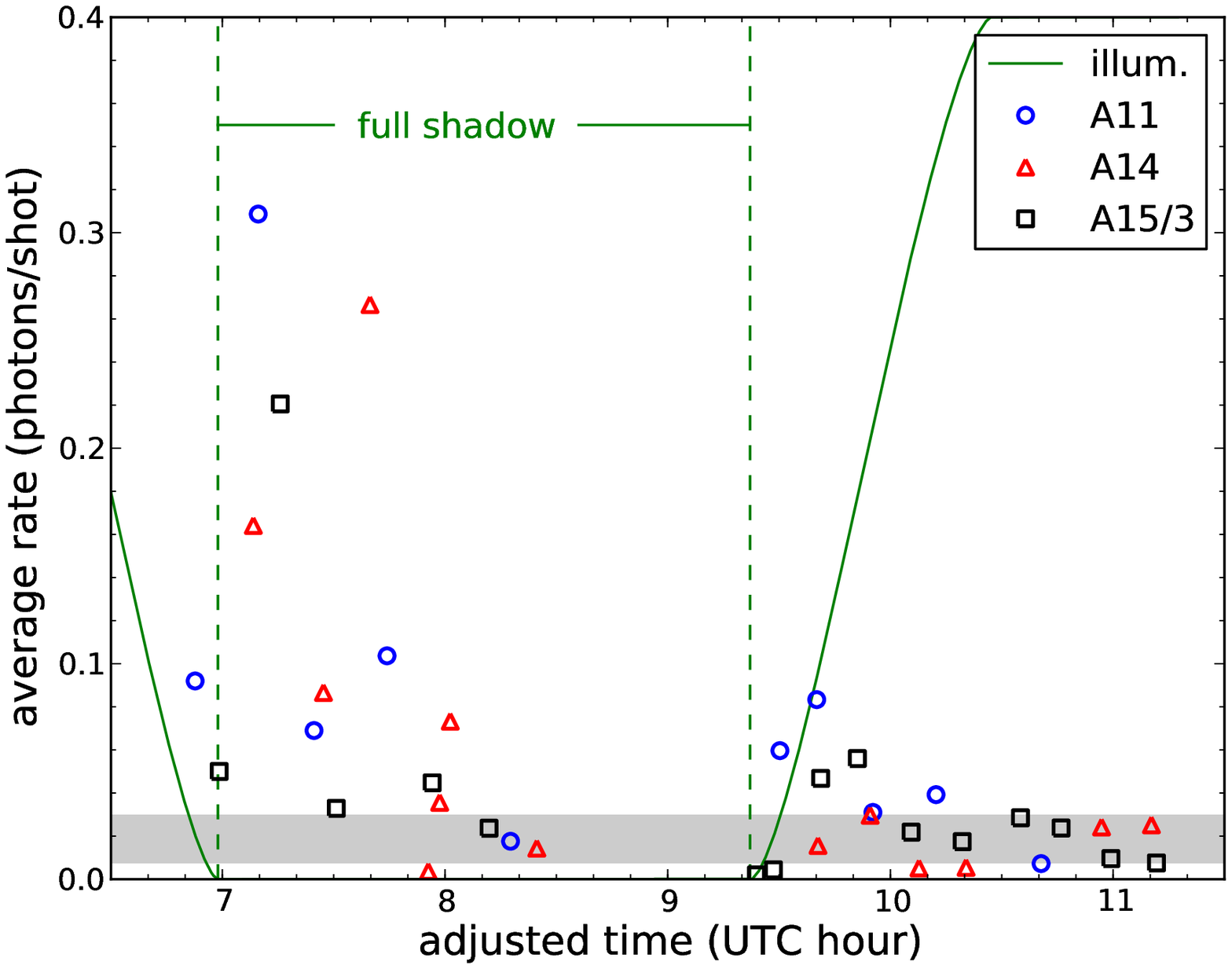}	
\hfill{}\includegraphics[width=88mm]{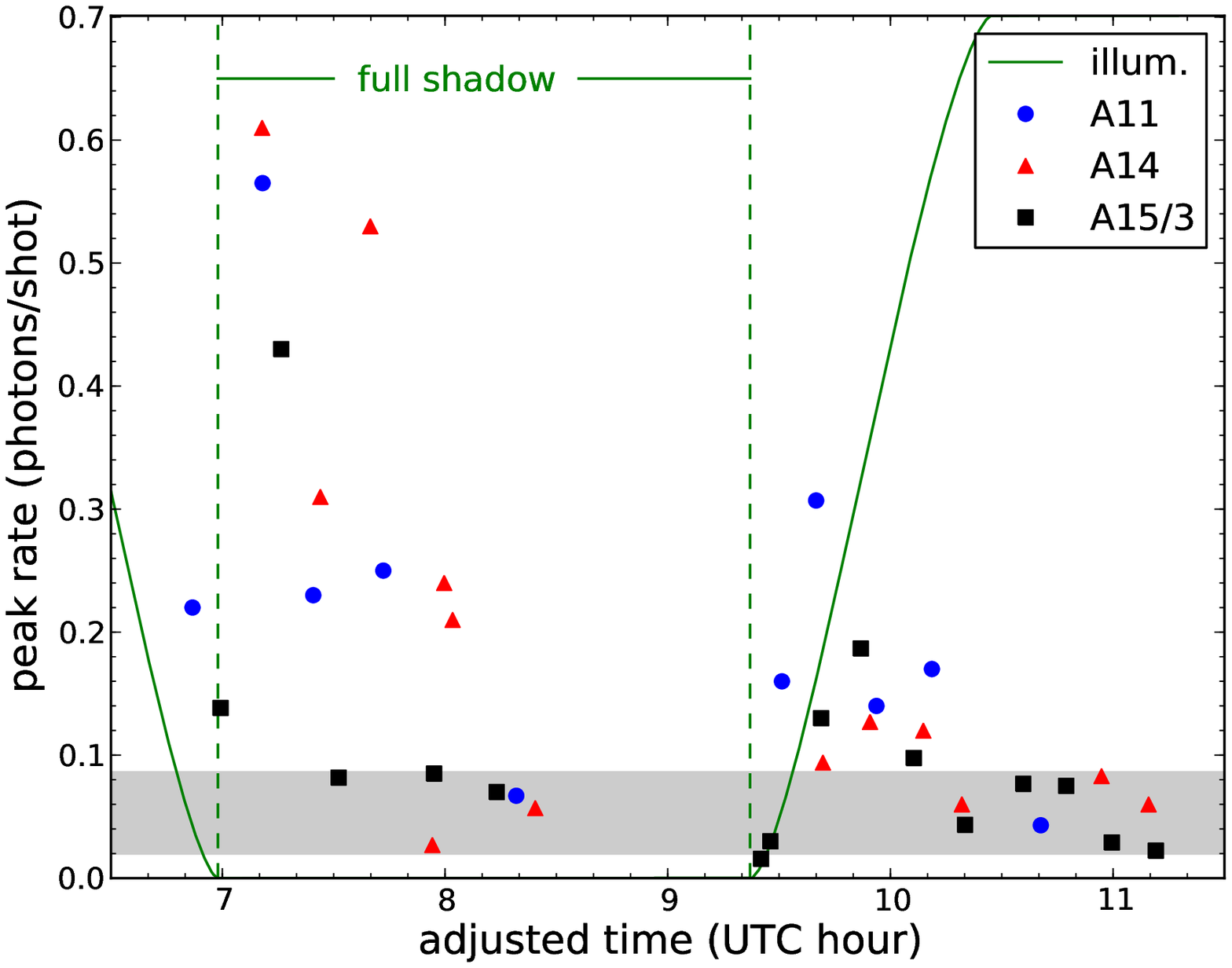}	
\caption{(color online) Return rates as a function of adjusted time,
synchronized with the Apollo~14 (earliest) illumination history. At left is
the average rate during each run, and at right is the peak rate. Times are
based on either the midpoint of the run or the time of the peak.  The
historic full-moon rates observed by APOLLO over the preceding five years
are indicated by the gray bands. The eclipse rates reach an order of
magnitude higher than historical results at full moon.  During the latter
part of totality, the signal became so weak that it was lost. Curves
indicating the state of illumination are shown in each plot, following
Fig.~\ref{fig:illum-curves}.\label{fig:adj-rates}}
\end{figure}

Adjusting the times to a common illumination history (as in Fig.~\ref{fig:ingress-egress})
reveals the trends a bit more clearly, as seen in Fig.~\ref{fig:adj-rates}.
All reflectors follow a similar evolution. Both peaks in performance
are seen to commence after the onset of shadow and re-illumination,
respectively, and display similar timescales. The second peak is about
half as strong as the first, although a systematic shift in atmospheric
transparency or seeing could be responsible for a change at this level.
Well after the end of the eclipse, the rates settle into a range consistent
with APOLLO's historically observed full-moon return rates over five
years of prior operation.

\subsection{Cloud Transparency Analysis}

\begin{figure}
\begin{center}\includegraphics[width=88mm]{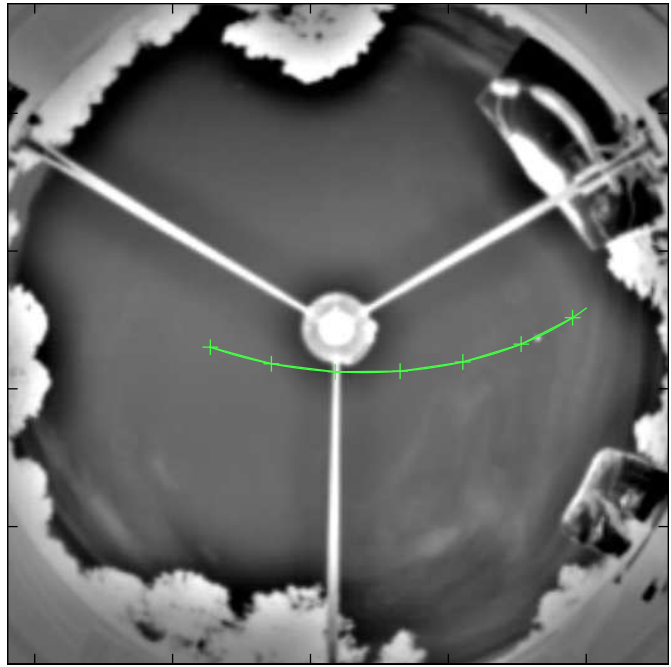}	
\end{center}
\caption{(color online) Example image from the infrared all-sky camera at
10:18:30 UTC. North is up, and east is to the left. The Moon's track is
superimposed, from 5:00 to 11:20, with plus symbols at the beginning of
each hour. The Moon is visible on the right side of the track, as well as
light clouds, which tended to move toward the ENE direction throughout the
observation. A cloud feature is seen to the west of the Moon in this image
that will result in 50\% opacity a few minutes later.\label{fig:irsc}}
\end{figure}

\begin{figure}
\begin{center}\includegraphics[width=88mm]{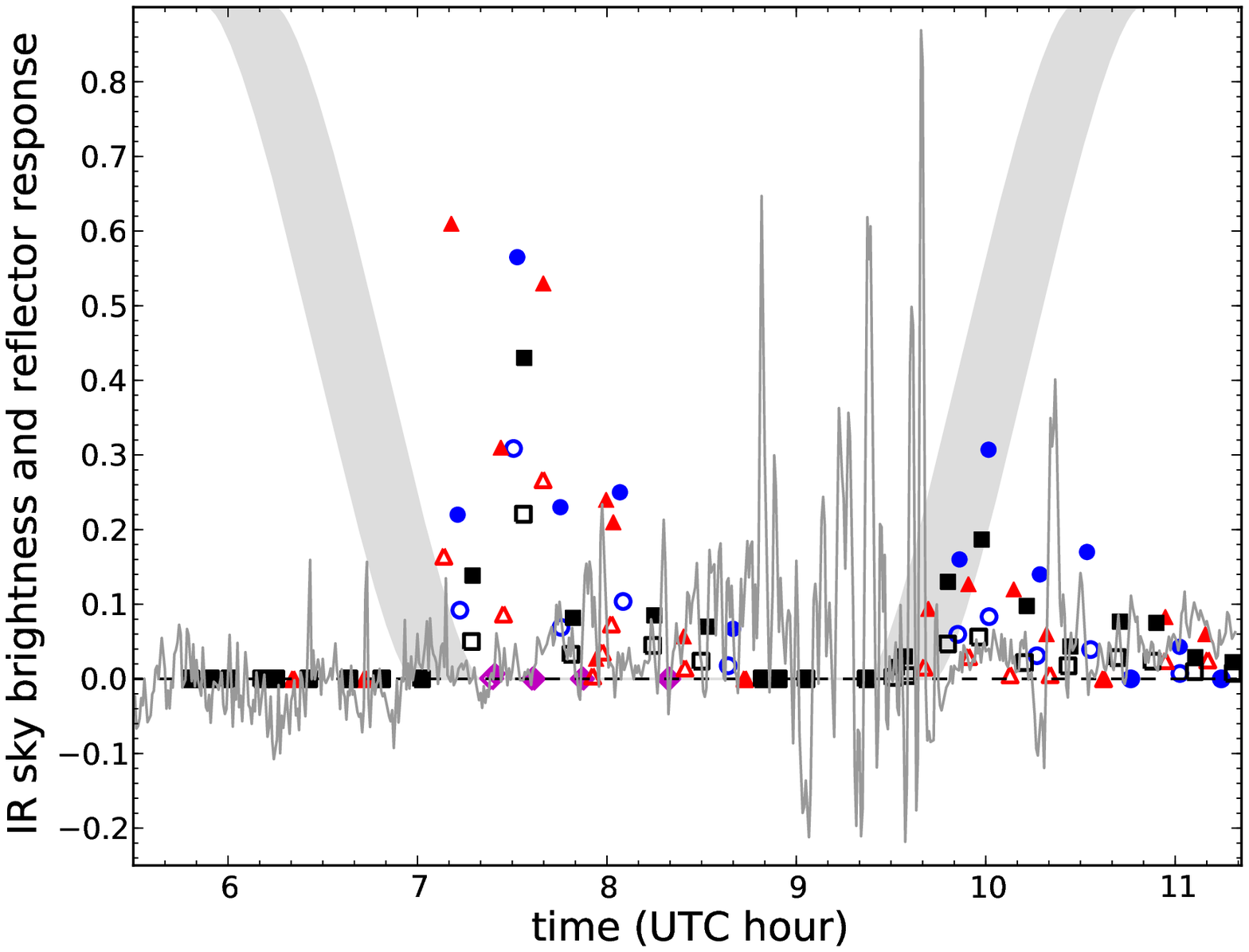}	
\end{center}
\caption{(color online) Observing history including estimated cloud
opacity.  The gray line represents infrared sky brightness at the position
around the Moon as a fraction of the highest brightness experienced at that
horizon-referenced location (ignoring the Moon itself) as clouds passed
during the night. Negative numbers reflect a localized auto-scaling of the
camera near cloud edges, and can be interpreted as representing clear
conditions. The spike at 10:20 is associated with a 50\% drop in the
infrared brightness of the Moon. Observations are overlaid, following the
conventions of Fig.~\ref{fig:raw-rates}, adding failed
observations---including those of Lunokhod~1 shown as diamonds. The light
gray band indicates the range of illumination curves for the three Apollo
reflectors, as shown in Fig.~\ref{fig:illum-curves}.  \label{fig:skyflux}}
\end{figure}

High-altitude clouds of variable thickness---mostly thin---were present
throughout the eclipse. The unprecedented signal rate at full phase is
convincing enough, but it is important to explore the extent to which
clouds may impact the signal profile. An all-sky infrared (IR) camera
working around 7--14~$\mu$m provides a record of sky conditions at the site
(see example in Fig.~\ref{fig:irsc}). By characterizing the range of
infrared brightness (proportional to cloud thickness) experienced at any
given location along the Moon's track during the period of observation, we
are able to establish the relative brightness of the sky around the Moon as
a function of time. Representing this brightness on a scale of zero (clear)
to one (maximum brightness experienced at that position during the
observation period; not necessarily opaque), we arrive at the erratic gray
curve in Fig.~\ref{fig:skyflux}. The spike at 10:20 was seen to diminish
the direct infrared flux from the Moon (which disappeared during the
eclipse) by a factor of two, while the smaller spike at 10:29 resulted in a
20\% reduction in flux. This reinforces the assertion that the clouds on
this night exhibit only modest opacity, although it should be borne in mind
that lunar ranging signals must traverse the atmosphere twice.

Also displayed in Fig.~\ref{fig:skyflux} is the observing history
following the conventions of Fig.~\ref{fig:raw-rates}, this time
adding failed attempts. The early and late periods are mostly clear,
while the highest opacity period from 8:20 to 9:40 coincides with
the loss of signal late in the eclipse. On its face, this would seem
to challenge our claim that the response pattern we observed stems
from intrinsic thermal problems in the corner cube prisms. But detailed
examination does not bear this out. 

Firstly, the initial rise and fall of signal transpired against a
backdrop of good conditions. Indeed, the clear-looking photograph
at left in Fig.~\ref{fig:photos} coincides with the worst conditions
during this period, while the picture at right---also evidently clear---corresponds
to a time when the lunar signal was already diminished. In other words,
the clouds in question are not very thick. Secondly, the cloudy periods
were always intermittent rather than steadily opaque. APOLLO observations
frequently succeed in intermittent conditions, experiencing peaks
and valleys in the response rate during runs lasting a few minutes.
Note in particular the failed attempt around 9:01 during momentarily
clear conditions. Lastly, signal recovery began before the most intense
cloudy period subsided at 9:40, after which the signal strength went
through another peak and decay during a period of good conditions. 

Even though the observing conditions were less than pristine on the
night of the eclipse, it should be strongly emphasized that the observations
are remarkable not only because the performance peaks happen to line
up with the times of thermal transition, as speculated should be the
case (e.g., Fig.~\ref{fig:cartoon}). More robust is the fact that
the returns from all three Apollo reflectors exceeded previous full-moon
return rates by an order of magnitude after the onset of totality,
eventually settling back to the historically observed rates after
the eclipse was over. While the observing conditions on the night
of the eclipse were decent, the presence of thin clouds would rule
out the prospect of beating previous records by an order of magnitude,
especially when simultaneously contending with a laser power that
was lower than the long-term normal. Variable conditions do not, therefore,
diminish the significance of the eclipse observations as an indicator
of thermal problems in the reflectors.

\subsection{Missing Lunokhod}

The lack of return from Lunokhod~1---normally comparable to the Apollo~11
and Apollo~14 response---is an interesting datapoint. The physically
larger, triangular-faced CCRs have a thermal time constant an order
of magnitude longer than the Apollo CCRs, and are not mounted in recesses
in an aluminum tray as are the Apollo CCRs. It is well-established
that the Lunokhod arrays suffer substantial performance degradation
under solar illumination. Having reflective (partially absorbing)
rear-surface coatings rather than operating via TIR may be partially
responsible. In general, the Lunokhod arrays are more susceptible to thermal
disruptions than the Apollo arrays, so the lack of returns from Lunokhod~1
during the eclipse is not entirely surprising.

\section{Discussion}

The present observations leave little doubt that the lunar reflectors
experience thermally induced performance degradation under solar illumination
at full moon---confirming the conjecture raised in \citet{dust}. Indeed,
it is rather remarkable that the return rates observed during this
one lunar eclipse approach the best levels observed by APOLLO at any
lunar phase over many years, despite the presence of high, thin clouds.
This is similar to the experience during the 1996 September 27 lunar
eclipse at the Observatoire de la C\^ote d'Azur \citep[OCA:][]{oca}:
immediately after ingress they recorded a return from the Apollo~11
reflector within a factor of two of their best-ever result at any
lunar phase, over decades of observation. Normally, OCA is unable
to acquire any LLR signal during full moon.

It is difficult to explain the observed double peak in signal strength
associated with eclipse ingress and egress without invoking a sign reversal
in the thermal gradient within the corner cube. We know that the gradient,
as defined in Section~\ref{sec:prediction}, must be negative during the
eclipse, as the front surface operates as a cooling surface radiating to
space. The implication is then that the gradient is positive during solar
illumination near full phase. Using the estimates and relations established
in Section~\ref{sec:thermal-model}, this puts the thermal input from dust
absorption greater than $\sim0.6$~W.  According to Eq.~\ref{eq:absorbed},
the dust covering fraction must be at least $f>0.25$ in order to exceed
0.6~W of absorption, which translates to a static (all phases) optical loss
of a factor of three.  The fact that the normal full-moon deficit is
substantial \citep{dust}, together with the foregoing observation that the
signal evolution during the eclipse is dramatically double-peaked, implies
a non-eclipsed solar absorption well in excess of 0.6~W---arguing for even
larger dust covering fractions and static losses. Indeed, as sketched by
Fig.~\ref{fig:frac-dep}, a covering fraction around $f\sim0.5$ is
consistent with: a) a factor of ten signal loss at all lunar phases; b) a
thermal gradient during solar illumination large enough to account for a
further order-of-magnitude signal loss at full moon; and c) the
double-peaked dynamic observed during eclipse.

The only piece that does not fit perfectly is the relative strengths
of the ingress and egress signal peaks. The fact that we were unable
to recover any signal during the late stages of totality suggests
that the negative gradient at that stage was stronger than the positive gradient
during full illumination. For this to be true---as would be the case
for the estimated net absorption of 0.4~W versus a $-0.6$~W cooling
rate in shadow---one expects the transition to be faster during ingress
than during egress. This, in turn, should lead to a weaker signal
peak during ingress than during egress, as depicted in Fig.~\ref{fig:cartoon}.
We see the opposite (Fig.~\ref{fig:adj-rates}). This discrepancy
is not overly concerning, however, in light of the fact that changing
conditions can easily produce variations in return strength on this
scale (factor of 2--3 level). Cloudier conditions combined with increased
wind shake and higher airmass in the latter half of the observations,
while not able to mimic the observed response dynamics and order-of-magnitude
variations, are sufficient to impact the observations in a manner
that is still consistent with the model.

Confirmation that a thermal gradient is responsible for the observed
full-moon deficit may also help explain why waning gibbous-phase laser
returns are often very strong \citep[as can be seen in Fig.~1 of][]{dust}.
Near full-moon phase, each reflector suffers a strong positive gradient
from dust absorption of solar illumination on the front surface. When
the sun angle becomes large enough, the front surface energy absorption
diminishes, and the gradient slowly becomes negative as stored thermal
energy is radiated to space. During this transition (gibbous phase),
the gradient should pass through zero and result in a stronger signal.
The timescales are much longer than those for the eclipse, and the
CCRs themselves are capable of much faster adjustment. But the balance
between solar forcing and radiative release is itself adjusting slowly,
carrying the CCR along a slow thermal trajectory.

One final comment on an item of general interest: the suggestion that the
lunar reflectors are covered in a $\sim50$\% coating of dust is not in
conflict with the survival of astronaut footpaths as dramatically revealed
in images from the Lunar Reconnaissance Orbiter Camera.  If the dust
buildup is a process that is linear in time, the present observation of
half-covered layer forming over $\sim 40$~yr suggests a deposition rate of $\sim
0.1$~$\mu$m~yr$^{-1}$---assuming typical grain sizes of $\sim 10\ \mu$m
\citep{grain-size}.  At this rate, it would take tens of thousands of years
to cover the tracks of the astronauts.  On the other hand, impact ejecta
rays are known to survive over 100~Myr timescales and may point to local
differences in surface composition with respect to regolith dust.

\section{Conclusion}

Estimates and observations provided in this paper support the idea
that dust accumulation could be responsible for several observed effects:
reduced signal at all phases; the full-moon signal deficit; and the
double-peaked signal during eclipse. The key idea is that when illuminated,
solar thermal absorption by a $\sim50$\% dust-covered CCR front surface
provides enough heat to support a several-degree thermal gradient
within the CCR, and this in turn is enough to destroy the central
irradiance of the diffraction pattern sent back to Earth. The double
peak in performance coincident with eclipse ingress and egress are
exactly what one would expect under this model. The timescales observed
for signal evolution are likewise realistic.

These results bear on the concept of optical or infrared observatories
on the lunar surface. On the timescale of decades, a layer of dust
may form on optical surfaces, compromising performance. It is unclear
how far from the surface the dust accumulation extends. A potential
hint is that the Lunokhod reflectors, situated roughly a meter off
the surface, appear to fare no better---and in the case of Lunokhod~2,
much worse---than the much lower-profile Apollo reflectors.  Results from
the recently launched  Lunar Atmosphere and Dust Environment Explorer
(LADEE) will likely shed significant light on dust transport mechanisms.

Also important is that next-generation lunar reflectors, such as those
being developed presently \citep{currie}, incorporate baffles or other
mechanisms to mitigate dust accumulation on critical surfaces. Independent
of the chief direction of transport, reducing the solid angle visible
from the corner cube front surface should help. As new reflectors
are developed and tested in simulated environmental conditions, it
could be highly informative to explore the role that dust plays in
the far-field performance of lunar corner cubes.

\paragraph{Acknowledgments}

We thank the Apache Point Observatory and the University of Washington for
providing a lengthy observing period during which to track the eclipse
progress. We are also grateful to Jack Dembicky for providing an inspiring
set of images of eclipse ranging operations. We thank Douglass Currie for
useful discussions, and Eric Michelsen and Jim Williams for useful
comments.  The apparatus used for these observations is associated with the
APOLLO collaboration, which also includes Eric Adelberger, James Battat, C.
D. Hoyle, Christopher Stubbs, and Erik Swanson.  APOLLO is supported by the
National Science Foundation, grant PHY-1068879, and NASA NNX-12AE96G, and
T. M. acknowledges support from the LUNAR consortium as part of the NASA
Lunar Science Institute (NNA09DB30A).  Results in this paper are based on
observations obtained with the Apache Point Observatory 3.5~m telescope,
which is owned and operated by the Astrophysical Research Consortium.

\end{document}